\documentclass{PoS}

\long \def \blockcomment #1\endcomment{}

\def\bq{\overline{q}}
\def\bj{\overline\j}
\def\bc{\overline\c}

\def\Dstag{D_{stag}}
\def\tx{\tilde{x}}
\def\ty{\tilde{y}}

\def\id{{\bf 1}}
\def\tD{\widetilde{D}}
\def\sl#1{\rlap{\hbox{$\mskip 2.5 mu /$}}#1}
\def\leqx{\,\raisebox{-1.0ex}{$\stackrel{\textstyle <}{\sim}$}\,}
\def\svev#1{\left\langle #1\right\rangle}
\def\LQCD{\Lambda_{QCD}}

\def\a{\alpha}

\def\c{\chi}

\def\e{\epsilon}                
\def\g{\gamma}

\def\j{\psi}

\def\m{\mu}

\def\x{\xi}

\def\D{\Delta}

\def\L{\Lambda}

\def\cd{{\cal D}}

\def\cf{{\cal F}}

\def\ck{{\cal K}}

\def\co{{\cal O}}

\def\ct{{\cal T}}
\def\cu{{\cal U}}
\def\cv{{\cal V}}

\def\inlinetilde{\lower0.8ex\hbox{$\,\widetilde{}\,$}}
\def\chpt{\raise0.4ex\hbox{$\chi$}PT}
\def\PQchpt{PQ\raise0.4ex\hbox{$\chi$}PT}
\def\Qchpt{Q\raise0.4ex\hbox{$\chi$}PT}
\def\schpt{S\raise0.4ex\hbox{$\chi$}PT}

\def\leftvec{{\raise1.5ex\hbox{$\leftarrow$}\kern-.85em}}
\def\half{{\scriptstyle \raise.2ex\hbox{${1\over2}$}}}
\def\threehalves{{\scriptstyle \raise.15ex\hbox{${3\over2}$}}}
\def\third{{\scriptstyle \raise.15ex\hbox{${1\over3}$}}}
\def\twothirds{{\scriptstyle \raise.15ex\hbox{${2\over3}$}}}
\def\fourth{{\scriptstyle \raise.15ex\hbox{${1\over4}$}}}

\def\gtwid{{\,\raise.3ex\hbox{$>$\kern-.75em\lower1ex\hbox{$\sim$}}\,}}
\def\ltwid{{\,\raise.3ex\hbox{$<$\kern-.75em\lower1ex\hbox{$\sim$}}\,}}

\def\eg{{\it e.g.},\ }
\def\et{{\it et al.}}

\def\cf{{\it cf.}\ }
\def\cO{{\cal O}}
\def\cL{{\cal L}}
\def\cM{{\cal M}}
\def\eqn#1{\label{eq:#1}}
\def\Equation#1{Equation~(\ref{eq:#1})}

\def\eq#1{Eq.~(\ref{eq:#1})}

\def\prd#1{Phys.\ Rev.\ D {\bf #1}}

\title{Regularizing QCD with staggered fermions and the fourth root trick}

\ShortTitle{Staggered fermions and the fourth root trick}

\author{Claude Bernard\\
Department of Physics, Washington University, Saint Louis, MO 63130, USA\\
E-mail: \email{cb@lump.wustl.edu}}
\author{Maarten Golterman\\
Department of Physics and Astronomy,
San Francisco State University,\\
San Francisco, CA 94132, USA\\
E-mail: \email{maarten@stars.sfsu.edu}}
\author{Yigal Shamir\\
School of Physics and Astronomy,
Raymond and Beverly Sackler Faculty of Exact Sciences,
Tel-Aviv University, Ramat~Aviv, 69978, Israel\\
E-mail: \email{shamir@post.tau.ac.il}}

\abstract{We investigate the properties of staggered-fermion lattice QCD in which the
fourth root of the fermion determinant is taken.  We show that this theory is
non-local at non-zero lattice spacing $a$, and that the non-locality is caused by the
breaking of taste symmetry at $a\ne 0$.   We then present a renormalization-group
based argument  that the theory restores taste symmetry in the continuum limit.
As a consequence the theory is local in that limit, and falls into
the correct
universality class.  Finally, we argue that the correct effective theory for the physics
of Goldstone bosons at $a\ne 0$ is given by staggered chiral perturbation theory
with the replica trick.
}

\FullConference{XXIVth International Symposium on Lattice Field Theory\\
                July 23-28, 2006\\
                Tucson, Arizona, USA}

\begin{document}

\newcommand\be{\begin{equation}}
\newcommand\ee{\end{equation}}
\newcommand\ba{\begin{eqnarray}}
\newcommand\ea{\end{eqnarray}}
\newcommand\tr{{\rm tr}}
\newcommand\Det{{\rm Det}}
\newcommand\ie{{\it i.e.}}
\newcommand\barp{{\overline{p}}}
\newcommand\hatp{{\hat{p}}}

\section{Introduction and overview}

Staggered fermions \cite {st} are commonly used for numerical computations in lattice QCD,
in which it is now becoming standard to include the fermion determinants for all three light
quark flavors, up, down, and strange.   They
 have an exact chiral $U(1)$ symmetry at zero quark mass \cite{ks}, implying that
the critical value of the bare quark mass for each quark flavor is known to be zero.
Furthermore, staggered fermions are numerically inexpensive, and the combination
of these two facts has made it possible to reach the chiral regime of light
pseudo-scalar Goldstone
boson (GB) masses essential for phenomenology.

In practice, a separate staggered fermion field is introduced for each physical flavor,
with a single-site mass term to describe their masses $m_u$, $m_d$ and  $m_s$.
However, each staggered fermion describes four degenerate quarks in the continuum
limit, commonly referred to as ``tastes,'' and one ends up with a theory with four up, four down,
and four strange quarks, with (for non-degenerate quark masses) a
$U(4)_u\times U(4)_d\times U(4)_s$ vector-like symmetry.  In order to remedy this
problem, one starts from the observation that if indeed each staggered fermion field
describes four degenerate tastes in the limit of vanishing lattice spacing $a$, one would expect that the
determinant of the staggered Dirac operator $D_{stag}$ factorizes as
\be
\label{factor}
\Det(D_{stag}(m))\sim\Det^4(D_{continuum}(m))\ ,\ \ \ \ \ a\to 0\ ,
\ee
where $m$ is the quark mass.
The idea is then to take the fourth root, $\Det^{1/4}(D_{stag})$ for each
flavor in the
generation of the gauge-field ensemble on which observables are computed.
At the diagrammatic level, this multiplies each sea-quark loop by $1/4$, thus
correcting for the too-many tastes per quark flavor present in the theory without the
fourth root.  Since
the staggered determinant is positive for any $m\ne 0$, and the continuum determinant
is (formally) positive for any $m>0$, one picks the positive fourth root.\footnote{A
negative
staggered quark mass can be made positive through a $U(1)_\epsilon$ rotation.}
Since the continuum determinant is only positive for $m>0$, the trick only works
for positive physical quark mass (see, however,
Refs.~\cite{bgss,Durr:2006ze}).

Obviously, the argument for this trick is heuristic.   Factorization as in
Eq.~(\ref{factor}) can never be exact:  first, the right-hand side is not well-defined,
and second, the ultraviolet (UV) eigenvalues of $D_{stag}$ will never form taste quartets.
But we expect that quartets of eigenvalues will form at physical scales for small
enough lattice spacing, while UV effects can be absorbed into a renormalization
of the gauge coupling and masses, and thus that factorization will effectively take place in the
continuum limit, validating the use of the fourth-root trick.  We note that this
expectation is based on the generally-assumed properties of the continuum limit
of \emph{unrooted} staggered  fermions (to which Eq.~(\ref{factor}) refers),
about which there is little doubt.  This
observation plays a key role in establishing the validity of regulating QCD with
rooted staggered fermions \cite{ys}.

We thus face the following set of questions, all of which need to be answered to
establish the validity of working with rooted staggered fermions:

\begin{itemize}
\item[1.]
The prescription for regulating QCD with staggered fermions and the
fourth-root trick is unambiguous and easy to implement.  The question is whether
this prescription is a regulator like any other or not.  We will argue in Sec.~2
that it is not, in the sense that the theory is non-local at $a\ne 0$ \cite{cmy}.

\item[2.]
Given this result, the next question is whether the continuum limit
can be taken, and whether that limit
is in the correct universality class.   In Sec.~3 we discuss a renormalization-group (RG) framework through
which to address this question \cite{ys}.  To motivate the use of the RG, we note that it
deals with two issues already mentioned:  it separates UV scales from
physical scales, and it makes it possible to define a $U(4)$ taste-invariant
theory  at some fixed coarse lattice spacing $a_c\ll \Lambda_{QCD}^{-1}$,
obtained by blocking the original staggered theory on a fine lattice with
spacing $a=a_f$ an infinite number of times (so that $a_f/a_c\to 0$).  In other words,
it helps in
defining the right-hand side of Eq.~(\ref{factor}).  Our arguments
 give strong evidence that indeed the
non-locality vanishes in the continuum limit and the
correct universality class is obtained using the fourth-root trick, \ie, that it is a
valid regulator.
All the key elements of the argument can be confirmed by perturbative
calculations and/or
concrete numerical tests.

\item[3.]
In view of the answers to questions  1 and 2, a third issue arises.
Lattice computations are performed at non-zero $a$, and they are thus
affected by the non-local nature of the regulator.   Even though the answer to
the second question implies that these unphysical effects go away in the
continuum limit, one needs to understand them in order to analyze and fit the
data generated at $a\ne 0$.   Here, this issue is addressed for the physics
of GBs.  In Sec.~4 we argue that the correct effective theory
describing  GBs is given by staggered chiral perturbation theory (\schpt)
plus the replica trick \cite{cb}.  At a pedestrian level, the replica trick refers
to the fact that sea-quark loops are corrected by a factor of four ``by hand,''
after identifying the quark-flow diagrams underlying a \schpt\ calculation.
The replica trick thus implements the fourth root at the level of the effective
theory.
\end{itemize}

Before embarking on the topics outlined above,
let us first consider the continuum limit in more detail.
If the continuum limit exists and is in the correct universality
class, it can be (formally) described by the path integral
(including sources for mesons)
\be
\label{cl}
Z_{continuum}(J)=\int {\cal D}{\cal U}\ e^{-S_g({\cal U})}\
\Det^{1/4}\Big((D({\cal U})+M)\otimes\id+J\Big)\ ,
\ee
where $S_g({\cal U})$ is the gauge action. The Dirac operator $D({\cal U})$
carries no taste index, and $\id$ is the $4\times 4$
identity matrix acting on the taste index.
The mass matrix is $M=diag(m_u,m_d,m_s)$,  with all
masses positive.
The source $J$ is a matrix in spin, flavor and taste space.
If we project the
complete set of correlation functions generated by this partition function onto
the taste-singlet set by setting $J={\tilde J}\otimes \id$,
the fourth root reduces to $\Det\big((D({\cal U})+M)+{\tilde J}\,\big)$,
and it  becomes obvious that the taste-singlet sector is the physical
sector, containing precisely the
correlation functions for unquenched three-flavor QCD.
This simple observation implies that no paradoxes based on symmetry arguments%
\footnote{as presented for instance in Ref.~\cite{mc}}
can arise \cite{bgss}.

Of course, many unphysical correlation
functions exist in the theory defined by Eq.~(\ref{cl}),
and they can be generated by using
a source $J$ that is not proportional to the identity in taste space.   However, because of the exact $SU(4)_{taste}$ vector
symmetry of the theory, flavor non-singlet but taste--singlet operators can be
related to taste non-singlet ones, for example,
\be
\label{example}\eqn{taste-rot}
{\overline u}\;\gamma_5\; d\
\longrightarrow_{\null_{\null_{\!\!\!\!\!\!\!\!\!\!\!\!\!\!\!\!\!\!\!\!\!\!\!\!\!\!\!\!\!\!\!\!SU(4)_{taste}}}}
\ {\overline u}\;\gamma_5\; \Xi\; d\ ,
\ee
where $\Xi$
is an $SU(4)_{taste}$ rotation acting (for example) on
the down quark.  If we choose $\Xi=\xi_5$, the operator on
the right-hand side corresponds to the exact GB related to $U(1)_{\epsilon}$
symmetry.  This is the pion field usually used in simulations.  Equation~(\ref{example})
tells us that this is equivalent to using the physical operator ${\bar u}\gamma_5 d$.
However, if one is interested in flavor-singlet physics, in which mixing
with gluonic states can occur, one must use taste-singlet operators, in accord
with the fact that gluons do not have taste.  Indeed,
 the trick of Eq.~(\ref{example}) will not work if both the quark and the anti-quark have the
same flavor. In cases with exact flavor symmetry, it may nevertheless
be possible to rotate a flavor-neutral interpolating field first into a flavor-charged
one, and then make taste rotations as in \eq{taste-rot}. For example, if $m_u=m_d$,
we can rotate a taste-singlet $\pi^0$ field into a taste-singlet $\pi^+$ field and
then use \eq{taste-rot} to relate it to a taste-nonsinglet  $\pi^+$ field. However,
this will not work if $m_u\not=m_d$, when a $\pi^0$ can mix with gluonic states.

\section{Non-locality at $a\ne 0$}

It is straightforward to establish the non-locality of the rooted theory at non-zero
lattice spacing.  We proceed by assuming the theory to be local, and derive a
contradiction \cite{cmy}.  Assume that a local Dirac operator $D$ exists such that (at $a\ne 0$)
\be
\label{assumption}
\Det^{1/4}(D_{stag})=\Det(D)\ {\rm exp}(-\delta S_{eff}/4)\ ,
\ee
with $\delta S_{eff}$ a local functional of the gauge field which does not
contribute to any long-distance effects (apart from a possible renormalization of
the gauge coupling).  Take the fourth power:
\be
\label{fourth}
\Det(D_{stag})=\Det(D_{4t})\ {\rm exp}(-\delta S_{eff})\ ,\ \ \ \ \ D_{4t}=D\otimes \id\ .
\ee
The operator $D_{4t}$ describes a theory with an exact $U(4)$ taste symmetry.
We may now compare the spectra of hadron multiplets in the two theories at non-zero
lattice spacing, considering
for instance the pions.  The theory defined by $D_{4t}$ has fifteen degenerate pions%
\footnote{Here we will use the word ``pion'' for any meson that becomes an exact
GB in the continuum and chiral limits.}
in the adjoint representation of $SU(4)$.  But the theory described by $D_{stag}$
is well known to have a spectrum of fifteen non-degenerate pions (with only one
``exact'' pion corresponding to $U(1)_\epsilon$ symmetry), due to the fact that
the staggered symmetry group is much smaller than $SU(4)$ \cite{gs}.  It follows that, contrary
to our assumption, $\delta S_{eff}$ has to know about long distance effects!
Thus the staggered theory with the fourth root is non-local.

While this concludes our basic argument,
it is instructive to discuss this result in more
detail.  In order to do this, we first go to the so-called taste basis, by defining \cite{ys04}
\be
\label{taste}
D^{-1}_{taste}=\alpha^{-1}+QD^{-1}_{stag}Q^\dagger\ ,
\ee
where $\alpha$ is a constant of order $1/a$, and $Q$ is a (gauge-covariant)
unitary matrix connecting the one-component and taste bases \cite{taste}.
The contact term $\alpha^{-1}$ is new,
but it has no effect on the long-distance physics.
Choosing $\alpha$
finite (instead of infinite, which would turn Eq.~(\ref{taste}) into the usual
basis transformation of Ref.~\cite{taste}) is advantageous
for setting up the RG framework in the
next section, as will become clear below.  With $0<\alpha<\infty$, we have that
\ba
\label{dets}
\Det(D_{stag})&=&\Det((\alpha G)^{-1})\;\Det(D_{taste})\ ,\\
(\alpha G)^{-1}&=&\frac{1}{\alpha} D_{stag}+Q^\dagger Q=\frac{1}{\alpha} D_{stag}+1\ .\nonumber
\ea
For $\alpha\sim 1/a$,  the effective action defined by
$\log{\Det((\alpha G)^{-1}))}$ is local, because $(\alpha G)^{-1}$ is a lattice Dirac
operator with a bare mass of order $1/a$.  This effective action vanishes as $\alpha\to\infty$.

We now may split $D_{taste}$ into taste-invariant and non-invariant parts:
\be
\label{split}
D_{taste}=D_1\otimes\id+\sum_A D_A\otimes \Xi_A\ ,
\ee
where the $\Xi_A$ are fifteen traceless hermitian matrices spanning the taste
space.  We thus have that
\be
\label{logdet}
\log{\Det(D_{taste})}=4\log{\Det(D_1)}+\log{\Det\Big(1+ D_1^{-1}\sum_A
  D_A\otimes \Xi_A\Big)}\ .
\ee
Both $D_1$ and the $D_A$ are local, but the effective action defined by the
last term on the right-hand side is not.  On the other hand, the theories
defined by $\log{\Det(D_{taste})}$ or $\log{\Det(D_1)}$ separately are obviously
local.\footnote{Note that when we say an
effective action is non-local, we mean that it cannot be written as
a fermion path integral with any local action.}
This makes it clear where the non-locality of the
rooted theory
comes from: it originates directly from the taste breaking present in the unrooted theory.
But it also makes it clear how it may go away in the continuum limit:
the $D_A$ are of order $a$, and thus constitute a set of irrelevant operators.
If no quantum effects change this observation, taste symmetry should be restored in the
continuum limit, and, \emph{mutatis mutandis} (as we will argue in the next section), the rooted theory should become local.    Equation~(\ref{logdet}) is an example of Eq.~(\ref{fourth}) with
$D=D_1$, but with $\delta S_{eff}$ non-local.

Working out $D_{taste}$ in the free theory in momentum space, we find that
\be
\label{free}
D_{taste}=\frac{\sum_\mu i[\gamma_\mu\otimes\id]\barp_\mu+
m+\frac{1}{\alpha}(\hatp^2+m^2)+
\frac{1}{2}\sum_\mu[\gamma_5\otimes\xi_\mu\xi_5]\hatp^2_\mu}
{1+\frac{2m}{\alpha}+\frac{1}{\alpha^2}(\hatp^2+m^2)}\ ,
\ee
in which $\barp_\mu\equiv\sin{p_\mu}$,
$\hatp_\mu=2\sin{(p_\mu/2)}$, and $\hatp^2\equiv\sum_\mu\hatp^2_\mu$.
The last term in the numerator of Eq.~(\ref{free}) is what removes
the fermion doublers in the usual taste-basis action \cite{taste} (\ie, for $\alpha=\infty$),
at the price of
breaking the taste symmetry explicitly.
The taste-invariant Dirac operator $D_1$ in
Eq.~(\ref{split}) is constructed by dropping that term.
An important observation is that $D_1$ also
has no doublers,
because of the Wilson-like term proportional to $1/\alpha\sim a$ in the numerator.
This is a prerequisite for having a taste-invariant theory
in the same universality class as the staggered theory,
making feasible the comparison of pairs of such theories in Sec.~3.

At this point, it is useful to reflect on the nature of the non-locality in the case at hand.%
\footnote{See also Ref.~\cite{sharpe}.}   In general, the tendency is to ``stay away''
from non-local field theories, whereas here we argue that in this case there is no
need to: the continuum limit is in the desired universality class ({\it cf.} next section), and the non-local
behavior at $a\ne 0$ can be understood in detail ({\it cf.} Sec.~4).  In the theory with the
fourth root, there are ``too many'' pions.  While we should end up with eight pions (in the
theory with three flavors),
before taking the fourth root the theory  has many more, because of the unphysical, extra taste
degree of freedom.  Of course, taking the fourth root is precisely intended to remove the
surplus of pions, by inserting the appropriate number of factors $1/4$ into
various terms making up the correlation functions describing the propagation of the
pions.
If all pions are members of exact taste multiplets,
this should work, as it would in the theory described by Eq.~(\ref{cl}).
The problem is that at non-zero lattice spacing taste symmetry is broken, so that
within one taste multiplet, the pion masses are (to leading order in \schpt) given by
\be
\label{pionmasses}
(m_\pi^A)^2=(m_\pi^{GB})^2+c^A a^2\Lambda_{QCD}^4\ ,
\ee
in which GB denotes the one exact Goldstone boson, and the $c^A$ are
numerical coefficients which do not vanish except when the taste index $A$ refers
to the exact GB.  This ``mismatch'' between masses shows up as violations of
unitarity, a manifestation of the non-locality of rooted
staggered fermions at
non-zero lattice spacing.  For explicit examples, see Refs.~\cite{PRELOVSEK,cdt,cb}.
We note that these arguments extend to other hadron multiplets as well;
we just focused on the pion sector because it contains the lightest excitations
in the theory.

One way to rephrase this observation is by noting that there are two independent
infrared  (IR) scales in the theory, one being the physical pion mass $m_\pi^{GB}$
governed by the quark mass, and the other $a\Lambda_{QCD}^2$,
which is generated by the taste splitting of low-lying eigenvalues of the
staggered Dirac operator.  In other words, there are two different scales
which control the IR behavior of the theory, a physical one and an unphysical
one.  As is often the case, the order in which IR scales are taken to zero matters,
and in this case it is clear that the right order is to first take the unphysical scale
$a\Lambda_{QCD}^2$ to zero if one wishes to study the chiral limit
\cite{cb2,bgss}.  Of course, taking $a\Lambda_{QCD}^2$ to zero is done by
taking the continuum limit.  In addition, it is clear that, in view of the
unitarity violations at $a\ne 0$, one should also take the continuum limit before
continuing the theory to Minkowski space.

\section{Renormalization-group analysis of (rooted) staggered fermions}

The (thus far formal) argument based on Eq.~(\ref{cl}) says that
rooted staggered fermions provide a valid regularization of QCD
if exact taste symmetry is recovered in the continuum limit.
We will now build an adequate non-perturbative framework where well-defined
statements can be made about the continuum limit.
Using this framework we will argue that under plausible, and testable,
assumptions, rooted staggered fermions
indeed provide a valid regularization of QCD.

\subsection{Strategy}

In Sec.~2 we discussed how taste-splittings manifest themselves
in physical observables.  We now take a step back and examine
taste-symmetry violations at the most fundamental level:
in the spectrum of the staggered Dirac operator.
On gauge-field configurations drawn from a
(rooted or unrooted) dynamical  ensemble,
what one expects to find is that the low-lying eigenvalues arrange
themselves nicely into almost-degenerate taste quartets \cite{evs,mu}.
But for larger eigenvalues the taste symmetry deteriorates
until, finally, at the cutoff,
taste degeneracy is completely lost.

If taste symmetry must be lost at the cutoff scale,
let's get rid of all cutoff-scale physics.
The way to eliminate cutoff-scale effects
is to apply renormalization-group (RG) block transformations.
With each blocking step the lattice spacing is doubled,
until an effective theory on a coarse lattice with spacing $a_c$ is reached.
As we approach the continuum limit
we  fix the QCD scale $\LQCD$ and the renormalized quark masses
by adjusting the bare parameters.
We fix the coarse-lattice spacing $a_c$ too:
each time we make an additional blocking step,
we simultaneously decrease the fine-lattice spacing $a_f$ by a factor of two.
The limit of infinitely many blocking steps implies
the usual continuum limit, $a_f \to 0$.
The coarse-lattice spacing provides a new, intermediate distance scale,
$a_f \ll a_c \ll \LQCD^{-1}$.
RG-blocking eliminates from the theory all
the fermion modes with eigenvalues (well) above the coarse-lattice cutoff
$1/a_c$. The remaining eigenmodes, those of the blocked Dirac operator,
all have eigenvalues that become vanishingly
small in units of the underlying cutoff $1/a_f$.
Because $a_c$ is kept fixed,
we expect that all these eigenvalues will {\it uniformly} show the
quartet structure required by taste symmetry.  In other words,
the blocked staggered Dirac operator will acquire the
taste-diagonal form of Eq.~(\ref{cl})
in the continuum limit, and its fourth root will correspond to
a local one-taste theory.

There are several reasons for choosing $a_c \ll \LQCD^{-1}$.
First, we take as criterion for
the theory in the continuum limit to be local  the requirement that the
coarse-lattice action be local on the scale $a_c$.
This only makes sense if $a_c \ll \LQCD^{-1}$.
We also want the complete set of coarse-lattice observables
to be rich enough to extract all of the QCD physics.
Again this requires that $1/a_c$ will be a high-energy scale
relative to the QCD scale.  A third reason will be encountered below.

The blocking framework constructed below is designed to
make the most out of our understanding of uncontroversial
lattice regularizations of QCD and, in particular, of unrooted
staggered fermions.  A fairly standard blocking framework is  set up
for ordinary (unrooted) staggered fermions in a form where the adaptations required for rooted
staggered fermions are minimal.

The novel element is the introduction of comparison blocked-lattice
theories with exact taste invariance.  At each blocking level,
a new theory with exact taste symmetry is constructed by simply dropping
the taste-breaking part of the blocked Dirac operator.
This is equivalent to dropping the rightmost, non-local term
of Eq.~(\ref{logdet}), except that it is done after a blocking process.
In the rooted theory, the result will be that
each of the so-obtained {\it reweighted}\
theories is a local one-taste theory.  Of course,
while all by itself RG-blocking leaves the physics invariant,
the reweighted theories constructed at different blocking levels
are different from one another, as well as
from the original (rooted or not) staggered theory.

Taste-breaking effects that survive blocking
become smaller and smaller with each additional blocking step.
As a result, at each blocking level,
the difference between a (blocked) staggered theory
and the corresponding reweighted theory gets smaller.
In the (continuum-)limit of infinitely many blocking steps,
the difference vanishes for every observable.\footnote{
  It is necessary to assume that $m\ne 0$ for all flavors.}
Because each reweighted theory is local
and belongs to the correct universality class,
the same applies to rooted staggered fermions in the continuum limit.

Clearly, this amounts to a set of highly
non-trivial claims which require a detailed justification.
Several key steps of the argument basically work in the same way for
the unrooted and the rooted staggered theories.
Note that for the ordinary (unrooted) theory, they lead to the uncontroversial
conclusion that the continuum-limit theory consists of
four degenerate quark species per staggered field.
The last crucial step of the argument is, however, more complicated
in the rooted theory.

\subsection{The RG blocking framework}

We first introduce our notation.  We will perform $n+1$ blocking
steps labeled $k=0,\ldots,n$. The $k^{\rm th}$ lattice spacing
is $a_k = 2^{k+1} a_f$ where as already mentioned
$a_f$ is the fine-lattice spacing and $a_c \equiv a_n$ is the coarse-lattice
spacing.  The $k=0$ step, already discussed in Sec.~2, is special.
It transforms the staggered field
from its usual one-component basis to a taste basis,
which is then retained in all subsequent blocking steps.
Thinning out of the fermion and gauge-field degrees of freedom occurs
at each step, except for the $k=0$ step where the fermions
are not thinned out but roughly speaking only undergo a change of basis.
In order to avoid overly cluttered notation we will consider
a singe-flavor theory in this section.
The generalization is trivial.

We first set up the blocking framework for
the ordinary (unrooted) staggered theory.\footnote{
  We assume a fixed, finite physical volume.
  By assumption all quarks are massive, and we expect no subtlety
  when taking the thermodynamical limit.
}
Blocking in the rooted theory will be introduced later.
The unrooted partition function is
\be
  Z = \int \cd\cu \cd\c \cd\bc\; \exp(-S_g -\bc \Dstag \c)\,,
\label{Zstag}
\ee
where
$\c(x),\bc(x)$ is a one-flavor staggered field,
and  $U_{\m,x}$ denotes the link variable.
The gauge field as a whole will be denoted $\cu=\{U_{\m,x}\}$.
Again $S_g=S_g(\cu)$ is the gauge action and $\Dstag=\Dstag(\cu)$
is the staggered Dirac operator.

The coordinates of the $k^{\rm th}$ step blocked lattice
will be denoted $\tx^{(k)}$. The fermion and anti-fermion fields on
that lattice are $\j^{(k)}_{\a i}(\tx^{(k)})$
and $\bj^{(k)}_{\a i}(\tx^{(k)})$ respectively.
The indices $\a$ and $i$, both ranging from one to four,
are the Dirac and the taste index respectively.
The blocked link variables will be denoted $V^{(k)}_{\m,\tx^{(k)}}$,
and the blocked gauge field
as a whole $\cv^{(k)}=\{V^{(k)}_{\m,\tx^{(k)}}\}$.

RG-blocking is performed by multiplying the integrand
of the path integral by one, written in a sophisticated form,
and then interchanging the order of integrations.
Fermion blocking is always done with a gaussian kernel that,
for $k\ge 1$, takes the explicit form
\be
  1 = \a_k^{-16N_k}
  \int \cd\j^{(k)} \cd\bj^{(k)}\;
  \exp\Big[
  \a_k
  \Big(\bj^{(k)} - \bj^{(k-1)} Q^{(k)\dagger}\Big)
     \Big(\j^{(k)} - Q^{(k)} \j^{(k-1)}\Big)
  \Big]\,.
\label{grass}
\ee
Here $\a_k$ is a blocking parameter of mass dimension one,
that is naturally taken to be $O(a_k^{-1})$,
and $N_k$ is the number of sites of the $k^{\rm th}$ lattice.
The blocking kernel $Q^{(k)}=Q^{(k)}(\cv^{(k-1)})$
is ultra-local and gauge covariant.
It defines a linear mapping from the sites of a $2^4$ hypercube
on the $(k-1)^{\rm th}$ lattice to the corresponding
single site of the $k^{\rm th}$ lattice.
Apart from the $k=0$ step (Sec.~2), $Q^{(k)}$ acts trivially on the Dirac
and taste indices.
For the gauge field we assume a conventional ultra-local blocking
kernel whose details are not needed.

We begin by introducing the kernels for the $n+1$ blocking steps.
However, we do not integrate over any gauge field yet,
for a reason that will become clear shortly. We do integrate
over all the fermion fields except those living on the coarse lattice.
This can be done in closed form because the fermion integrals are gaussian.
The result is
\ba
  Z &=& \int \cd\cu\, \cd\cv^{(0)}\, \cd\cv^{(1)} \cdots \cd\cv^{(n)} \;
  \exp\bigg(-S_g - \sum_{k=0}^n \ck_B^{(k)}
  - \sum_{k=0}^n S_{eff}^{(k)}\bigg)
\nonumber\\
    && \hspace{5ex}\times
  \int d\j^{(n)} d\bj^{(n)}\; \exp\Big( -\bj^{(n)} D_n\, \j^{(n)} \Big) \,.
\label{gauss}
\ea
Here $\ck_B^{(0)}=\ck_B^{(0)}(\cv^{(0)},\cu)$ and
$\ck_B^{(k)}=\ck_B^{(k)}(\cv^{(k)},\cv^{(k-1)})$, $1\le k\le n$,
are the gauge-field blocking kernels.
The UV fermion modes
that have been integrated out at the $k^{\rm th}$ step
give rise to the {\it effective action}
\be
  S_{eff}^{(k)} = \log \det(G_k) \,.
\label{SeffG}
\ee
The operators $D_k$ and $G_k^{-1}$ may be constructed iteratively
(for the $k=0$ step see Sec.~2)
\ba
  D_k &=&
  \a_k -\a_k^2\, Q^{(k)} G_k Q^{(k)\dagger} \,,
\label{DG}
\\
  G_k^{-1} &=& D_{k-1} + \a_k Q^{(k)\dag} Q^{(k)}\,.
\label{Gk}
\ea

Let us briefly summarize the important properties of these
operators in the free theory \cite{ys04}.
The operator $D_k$ is gap-less for $m=0$, with its low-lying eigenmodes
coming from $p\sim 0$.  In Eq.~(\ref{Gk}) the addition
of the blocking-kernel term lifts the small eigenvalues
and creates an $O(\a_k)=O(1/a_k)$ gap in the spectrum of $G_k^{-1}$.
One can then prove iteratively that $D_k$, $G_k^{-1}$, and $G_k$ are all local
operators \cite{cmp}.  By this we mean that
their matrix elements connecting sites $\tx^{(k)}$ and $\ty^{(k)}$
vanish exponentially with the separation, with an $O(a_k)$ decay length.

Expanding the free, coarse-lattice Dirac operator $D_n$
to second order in $p$ and $m$ gives
\be
  D_n(p)
  =
  m + i[\sl{p}\otimes \id]
   + a_f \sum_\m [\g_5 \otimes \x_5\x_\m]\, p_\m^2
  - R_n \big(m+i[\sl{p}\otimes \id] \big)^2 + \cdots \,.
\label{pexp}
\ee
where $R_n = \sum_{k=0}^n (16)^{n-k}/\a_k$ is $O(a_c)$.
The shown taste-breaking term originates from the last term
in the numerator of Eq.~(\ref{free}).  Because $|p|\; \leqx\; a_c^{-1}$,
this irrelevant term is $O(a_f/a_c^2)$, which is
just the scaling required by its engineering dimension.
In the limit $n\to\infty$ the blocked Dirac operator becomes taste-diagonal.
For $m=0$, the operator $D_n$ satisfies a Ginsparg-Wilson (GW) relation.
Thus, through RG blocking, the taste-violating mechanism
for removing the doublers is gradually taken over by the taste-conserving
{\it and}\
chiral symmetry conserving GW mechanism.

In both the free and the interacting theories
the inverse blocked Dirac operator satisfies\footnote{
  For the $k=0$ step, Eq.~(\ref{Dk}) reduces to Eq.~(\ref{taste}).
}
\be
  D_k^{-1} = \a_k^{-1} + Q^{(k)} D_{k-1}^{-1} Q^{(k)\dagger} \,.
\label{Dk}
\ee
Using Eq.~(\ref{Dk}) recursively we may express the coarse-lattice
fermion propagator as the fine-lattice propagator between smeared sources
defined by the product of the blocking kernels.\footnote{
  This is true up to the contact term in the propagator, which vanishes for
  $\tx^{(k)} \ne \ty^{(k)}$.
}
Equation (\ref{Dk}) is in fact a special case of a completely general
pattern.  Considering any operator $\co^{(c)}$
constructed from the coarse-lattice
fields, we may ``undo''
the blocking by performing the integration
over all the blocked (gauge-field and fermion) fields we have introduced,
resulting in an operator $\co^{(f)}$ that depends only on the fine-lattice
fields.  This operation defines a {\it pull-back}\
mapping  $\ct^{(f,c)}: \co^{(c)} \to \co^{(f)}$.
The pull-back mapping is ultra-local (because the blocking kernels are)
and it preserves expectation values:
$\svev{\ct^{(f,c)} \co^{(c)}}_f = \svev{\co^{(c)}}_c$.

The existence of the pull-back mapping means that every coarse-lattice
observable is at the same time also a fine-lattice observable.
This ``kinematical'' feature has an important dynamical consequence:
The coarse-lattice observable must preserve the constraints
that follow from all the fine-lattice symmetries, even
if (as it actually happens) some of these symmetries are not
manifestly preserved by the blocking (for more details, see Ref.~\cite{ys}).

We next introduce a family of
comparison theories as follows.  First, as in Eq.~(\ref{split})
we split the blocked Dirac operator
into its taste-invariant and taste-breaking parts
\be
  D_n = \tD_n \otimes \id + \D_n \,,
\label{splitn}
\ee
where $\tD_n$ carries no taste index, and $\D_n$ is traceless
on the taste index.  The replacement $\D_n \to t\D_n$,
where $0\le t\le 1$, then allows us to gradually ``turn off''
all the taste breaking.  For $t=0$, exact taste symmetry is restored by hand.
Performing this at the level of the (blocked) partition function
generates a family of {\it interpolating}\
theories
\ba
  Z_n^{inter}(t)
  &=& \int \cd\cu\, \cd\cv^{(0)}\, \cd\cv^{(1)} \cdots \cd\cv^{(n)} \;
  \exp\bigg(-S_g - \sum_{k=0}^n \ck_B^{(k)} - \sum_{k=0}^n S_{eff}^{(k)}\bigg)
\nonumber\\
    && \hspace{5ex}\times
  \int d\j^{(n)} d\bj^{(n)}\;   \exp\Big[ -\bj^{(n)}
  \Big( \tD_n \otimes \id + t\D_n \Big)\, \j^{(n)} \Big] \,.
\label{gausst}
\ea
In the next subsection we will argue that
these interpolating theories are all local.
They clearly belong to the same universality class
as the original staggered theory.

We now come to the rooted staggered theory defined by
\be
  Z^{root} = \int \cd\cu \;  \exp(-S_g)\;
  \Det^{1/4}(\Dstag)\,.
\label{Zroot}
\ee
The gaussian fermion-blocking transformations readily
lead to identities for the fermion determinant that hold for
any given set of values of the (original and blocked) gauge fields.
This allows us to write down the analogue
of Eq.~(\ref{gauss}) for the rooted theory
\ba
  Z^{root} = \int \cd\cu\, \cd\cv^{(0)}\, \cd\cv^{(1)} \cdots \cd\cv^{(n)} \;
  \exp\bigg(-S_g - \sum_{k=0}^n \ck_B^{(k)}
  - \frac{1}{4} \sum_{k=0}^n S_{eff}^{(k)}\bigg)
  \Det^{1/4}\big(D_n\big) \,. \qquad
\label{Z4n}
\ea
Of course, as already noted in Eq.~(\ref{logdet}),
after $n$-step blocking we still cannot express the rooted determinant
as a local path integral.  This is made possible only after dropping
the taste-breaking part $\D_n$ completely.
We then arrive at a {\it reweighted}\
theory
\ba
  Z^{reweigh}_n
  &=& \int \cd\cu\, \cd\cv^{(0)}\, \cd\cv^{(1)} \cdots \cd\cv^{(n)} \;
  \exp\bigg(-S_g - \sum_{k=0}^n \ck_B^{(k)}
  - \frac{1}{4}  \sum_{k=0}^n S_{eff}^{(k)}\bigg)
\nonumber\\
    && \hspace{5ex}\times
  \int dq^{(n)} d\bq^{(n)}\; \exp\Big( -\bq^{(n)} \tD_n\, q^{(n)} \Big) \,,
\label{Z4ninv}
\ea
where we used $\Det^{1/4}(\tD_n\otimes\id)=\Det(\tD_n)$.
The coarse-lattice quark fields $q^{(n)}$, $\bq^{(n)}$ carry no taste index.
In the next subsection, we will argue that the reweighted theories
$Z^{reweigh}_n$ are local too,
and that they fall into the right universality class.
(This puts on a solid basis the observations made below Eq.~(\ref{cl}).)
For rooted (unrooted) staggered fermions,
we will show that each observable of the $n^{\rm th}$ reweighted
(interpolated) theory has the same $n\to\infty$ limit as the corresponding
observable of the blocked staggered theory.

\subsection{Validity of rooted staggered fermions in the continuum limit}

Ordinary staggered fermions define a renormalizable lattice theory.
The renormalizability of the ordinary staggered theory
has not yet been established to all orders.
Yet there is no real reason to doubt
that all-orders renormalizability holds in the unrooted theory.
As shown in Ref.~\cite{gs}, any relevant or marginal term
not already present in the original staggered action
is forbidden by its symmetries, and thus will not be induced by
loop corrections.

Renormalizability readily generalizes to the rooted theory.
This was first observed in Ref.~\cite{PQ}.
For a recent detailed discussion, see Ref.~\cite{sharpe}.
Consider first a theory with $n_R$
copies, or replicas, of identical staggered fields.
The counterterms at any order are polynomials in (the integer) $n_R$.  Now,
in perturbation theory, rooting amounts to multiplying each staggered-fermion
loop by $1/4$. Therefore the same counterterms will suffice to renormalize
the rooted theory when we substitute $n_R \to 1/4$.
While locality is lost with rooting, renormalizability is saved.
The theory remains tightly constrained by power counting and symmetries.

Our third motivation for choosing
the coarse-lattice spacing to satisfy $a_c \ll \L^{-1}$,
is that QCD is weakly coupled at short distances.
The existence of a weak-coupling regime
depends on renormalizability, but it does {\it not}\
necessarily require locality.\footnote{
  See Ref.~\cite{sharpe} for a discussion of non-local but renormalizable
  two-dimensional field theories.
}
Any four-dimensional lattice gauge theory whose Boltzmann weight contains
a fermion determinant raised to a real positive power $n_R$
will have a weak-coupling regime,
so long as the one-loop beta function (that depends linearly on $n_R$)
remains asymptotically free.
All we need for the next step is that
the blocking process in its entirety is taking
place in a weak-coupling regime, in both the unrooted and rooted theories.

Next, we argue that the blocked Dirac operator $D_n$
and the effective action $S_{eff}^{(k)}$ (obtained by removing short-distance
fermion modes) are local, and that this is true on both the unrooted
and rooted ensembles.
Functional derivatives of $S_{eff}^{(k)}$ depend on $H_k^{-1}$
and on (derivatives of) $H_k$,
where  the
operator $H_k = [\g_5 \otimes \x_5] G_k^{-1}$ is hermitian \cite{cmy}.  In view of
the discussion below Eq.~(\ref{Gk}),
the necessary and sufficient condition for $S_{eff}^{(k)}$ to be local
is that both $H_k$ and its inverse be local operators on the $k^{\rm th}$
lattice scale, in the interacting theory too.
Unlike in the free theory, however, we cannot rule out the possibility that $H_k$ has {\it no}\
gap in the interacting theory, \ie,
that $H_k$ may have arbitrarily small eigenvalues.
But for the locality of $H_k^{-1}$ small eigenvalues are
harmless if the eigenmodes are exponentially localized
on the corresponding lattice scale.  Therefore, a sufficient condition
for the locality of $H_k^{-1}$, and iteratively for the locality
of $D_k$ and $H_{k+1}$, is that the {\it mobility edge}\
of $H_k$ be $O(\a_k)=O(1/a_k)$.  Eigenmodes with eigenvalues above the mobility edge
are, by definition, extended.  But as long as the extended-modes spectrum
does not reach down to zero, the inverse of $H_k$ will stay local.

A recent study of the mobility edge
of the Wilson operator in the super-critical region
reveals a mobility edge remarkably close to the free-theory gap
for moderate values of the bare coupling \cite{mob}.
By interpolation, the mobility edge must be even closer to
the free-theory gap in a (really) weak-coupling regime.
Intuitively this can be understood as follows.
Eigenvalues far below the free-theory gap arise in the presence
of ``dislocations'' in the gauge field.
But the pure Yang-Mills action on the $k^{\rm th}$ lattice
strongly suppresses dislocations provided that
the running coupling $g_r(a_k)$ is very small,
in which case the effect of any fermion determinant is subdominant
(the large, UV divergent part of
the fermion determinant is absorbed into the renormalization of $g_r(a_k)$).
This works in any
renormalizable and asymptotically free (but not necessarily local) theory,
in particular for both the ordinary
and the rooted staggered ensembles.  Having said that,
since none of the operators $H_k$ has been studied numerically to date,
it is clearly important to confirm this part of the argument.

If $D_n$ is local then, trivially, $\tD_n$ and $\D_n$ are separately
local, \cf Eq.~(\ref{splitn}).
Integrating over all the gauge-fields except $\cv^{(n)}$
in Eqs.~(\ref{gausst}) and (\ref{Z4ninv}), we obtain
coarse-lattice theories which are {\it local}.
In particular, the reweighted theories (\ref{Z4ninv}) are one-taste theories
that belong to the appropriate universality class.

What remains to be established is that the sequence of reweighted
(interpolating) theories has the same $n\to\infty$ limit
as the blocked rooted (unrooted) theory.
We now derive this result assuming that,
in both the interpolating theories (\ref{gausst})
and the reweighted theories (\ref{Z4ninv}),
the following scaling relations are valid in an {\it ensemble average}\
sense:
\ba
  \left\| D_n^{-1} \right\| & \leqx & \frac{1}{m_r(a_c)} \,,
\label{mscale}
\\
  \left\| \D_n \right\|  & \leqx &  \frac{a_f}{a_c^2}
  = {2^{-(n+1)}\over a_c}\,,
\label{irrscale}
\ea
where $m_r(a_c) > 0$ is the renormalized quark mass.
The scaling law (\ref{irrscale}) of the taste-breaking part $\D_n$
of the Dirac operator neglects logarithmic corrections.
It is the anticipated scaling based on the smallest
engineering dimension of an irrelevant operator, namely five.\footnote{
  See Ref.~\cite{FM} for a first direct study of the scaling of $\D_n$.}

In the rest of the argument we focus on the rooted theory.
We compare the rooted
and the reweighted theories at the $n^{\rm th}$ blocking level  starting from
one-taste operators constructed from the coarse-lattice fields of the
reweighted theory, \ie, operators of the form $\co^{(n)}=\co^{(n)}(q^{(n)},\bq^{(n)},\cv^{(n)})$.
For such operators, we reconstruct  rooted observables,
using that  un-normalized expectation values
satisfy\footnote{Eq.~(\ref{resum})
only relates a subset of the observables in the rooted theory --- those in the
physical subspace --- to those of the reweighted theory. If we want to relate
all rooted observables, we can (a) leave the reweighted theory
as a theory of four rooted, but equivalent, tastes, as was done
for the continuum theory in Eq.~(\ref{cl}), or (b)
write the reweighted theory as a theory of four unrooted tastes and three ghost
tastes, as in Appendix B of Ref.~\cite{sharpe}.
}
\ba
  \svev{\co^{(n)}}_n^{root}
  &=&
  \svev{ \co^{(n)} \; \exp \left[{1\over 4}\, \tr \log
  \Big( 1 + \D_n\, [\tD_n \otimes \id]^{-1} \Big) \right]\, }_n^{reweigh}
\nonumber
\\
  &=& \rule{0ex}{4ex}
  \svev{\co^{(n)}}_n^{reweigh} \left( 1 + O(\e_n^2)\right) \,,
\label{resum}
\ea
where we have used the taste-tracelessness of $\D_n$, and where,
using Eqs.~(\ref{mscale}) and (\ref{irrscale}),
\be
  \e_n \equiv \Big\| D_{inv,n}^{-1} \Big\|\, \Big\| \D_n \Big\|
  \;\leqx\; {a_f\over a_c^2\, m_r(a_c)}
    = {2^{-(n+1)}\over a_c\, m_r(a_c)}\,.
\label{eps}
\ee
The (anticipated) scaling of $\D_n$ thus implies that $\e_n\to 0$
for $n\to\infty$.  The observation here is that, first, after sufficiently
many blocking steps the expansion of the logarithm is convergent,
and second, that in the limit  $n\to\infty$
the expectation value of $\co^{(n)}$ is the same for the rooted
and for the reweighted
theories.\footnote{Strictly speaking, one should take into account the fact that
the valence propagators coming from $\co^{(n)}$ are also slightly different on
the two sides of Eq.~(\ref{resum}): The rooted theory keeps taste-violating
terms in these propagators but the reweighted theory does not.  The difference
is also $O(\e_n^2)$ and thus not important to the argument.}
This is precisely what is
needed for the validity of the rooted theory in the continuum limit!
A corollary is that the continuum-limit theory
enjoys every exact symmetry that exists in the (rooted) staggered and/or
in the reweighted theory.

\subsection{Scaling}

Let us now take a closer look at the scaling laws.
A standard perturbative treatment
in either the unrooted or the rooted theory
would predict all the scaling laws we have used:
for the running of $g_r$ and $m_r$,
and for the scaling of taste-breaking irrelevant terms.
But this falls short of what we need.

Scaling laws for irrelevant operators are usually derived
in a Symanzik effective-action context \cite{symanzik},
where one imagines integrating out
the short-distance fluctuations of all the lattice fields.
Here, in contrast, the scaling laws
must hold in Eq.~(\ref{Z4ninv}) (or in Eq.~(\ref{gausst})),
where a whole ``tower'' of gauge fields is still present.
How do we know that any scaling law still applies?
We have only assumed that the scaling laws (\ref{mscale})
and (\ref{irrscale}) hold in an ensemble average sense.
Furthermore, we have only assumed these scaling laws within the
reweighted theories, and these have a local
path-integral representation.
We do {\it not}\
rely on the validity of any scaling laws for irrelevant operators
directly in the rooted theory.

Operationally,
this means that the scaling laws
must hold for the expectation values of operators
constructed from the coarse-lattice fields of the reweighted theory.
Before we can compute any scaling law, we must first set up
perturbation theory.
Usually  lattice perturbation theory begins with the expansion
of the link variables as $U_{\mu,x}=\exp(iga A_{\mu,x})$.
In a reweighted theory
a similar expansion will have to be applied to the tower of gauge fields
$\cu,\cv^{(0)},\cv^{(1)},\ldots,\cv^{(n)}$ simultaneously.
With this, the perturbative expansion
can in principle be derived directly from Eq.~(\ref{Z4ninv}),
because the closed-form expressions for $D_n$, $\tD_n$
and $S_{eff}^{(k)}$ as functionals of all the gauge fields are known.
One would then proceed to calculate correlation functions with coarse-lattice fields
only on the external legs.  Because
only coarse-lattice gauge fields occur on any external leg,
we may imagine that the integrations
over $\cu,\cv^{(0)},\cv^{(1)},\ldots,\cv^{(n-1)}$ are always done
before the integrating over $\cv^{(n)}$.
Since, in addition, all the external momenta are of order $1/a_c$,
this brings us closer to a standard RG setup
where coarse-lattice observables are computed using a conventional
(but complicated) coarse-lattice action (that includes for example
multi-fermion interactions).  As we explain below, some differences
still remain.  But first we turn to a conceptual question.

There is a
crucial difference between the running of $g_r$ and $m_r$,
and the anticipated scaling of an irrelevant operator.
The running of relevant and marginal parameters
originates directly from the short-distance divergences of the theory.
Because power-counting and renormalizability
survive rooting, we have no reason to doubt the validity
of rooted perturbation theory for these specific scaling laws.
This is especially true given that
only general features of the running coupling and mass parameters
are needed, but not any details.

In contrast, the anticipated scaling
of any irrelevant operator (which  does not mix with relevant or
marginal operators) implies
the vanishing of that operator in the continuum limit.
That something should vanish in a certain limit
is a more delicate claim.
In local theories we have no reason to doubt the perturbative prediction,
which amounts to the assumption that no non-perturbative effects
interfere with the vanishing of all amplitudes with an insertion
of the irrelevant operator.  But, the question is,
how can we be sure that the non-locality
of the rooted theory does not modify the scaling
of irrelevant operators in undesirable
ways not captured by perturbation theory?
Once again, our solution is to rely only on the predictions
of perturbation theory in the {\it local}\
reweighted theories.  Because the non-locality has been eliminated,
we see no reason to trust reweighted perturbation theory
any less than perturbation theory for, say, unrooted staggered fermions.
This applies in particular to the scaling of (taste breaking)
irrelevant operators within the reweighted rooted theory, which, we expect,
would give rise to Eq.~(\ref{irrscale}).

In  the unrooted theory, an analogous analysis
starts off with the reweighted unrooted theory, \ie,
the $t=0$ interpolating theory (\cf Eq.~(\ref{gausst})),
and ends with the uncontroversial conclusion
that exact taste symmetry is recovered in the staggered theory
in the continuum limit. But, in the unrooted staggered theory
we could instead rely directly on unrooted perturbation theory
for the scaling of $\D_n$, because the unrooted theory itself is local.

Within reweighted unrooted perturbation theory,
the scaling (\ref{irrscale}) should hold as we go down from $1/a_f$
to $1/a_c$.\footnote{
  Note that this does {\it not}\
  amount to a standard blocking process within the reweighted theory.
}
This, in turn, should result from the similar symmetry patterns
of the staggered and the reweighted theories, for $n$ large enough.
On the staggered side, both taste symmetry and
(softly broken) chiral symmetry in the continuum limit
are secured by the lattice symmetries.
On the reweighted side,
we have exact taste symmetry by construction.
But if the continuum limit is to come out right, then the reweighted theory
should in addition have an approximate chiral symmetry (namely,
the additive renormalization of the taste-singlet mass term
goes to zero with increasing $n$).   Indeed, since the $t=0$ and
$t=1$ theories in Eq. ~(\ref{gausst}) are connected by a convergent
expansion, this must be the case.

We can summarize our line of reasoning as follows.  First, we claim that
the reweighted version of the unrooted staggered theory is local,
and that $\D_n$ scales as expected in that theory.   If one accepts the
validity of the unrooted staggered theory this claim can be considered
``safe.''  Then, proceeding from the unrooted reweighted theory, which
has four tastes and exact $U(4)$ taste symmetry, we can consider
the theory in which we take the fourth root of the fermion determinant,
and obtain a local one-taste theory in which
$\D_n$ still scales as an irrelevant operator.
Of course, the  gauge ensemble is different for the reweighted versions
of the rooted and unrooted theories. But
$\D_n$ should remain irrelevant as we move from the unrooted reweighted to
the rooted reweighted case since perturbation theory should be
equally trustworthy for both local theories.
Note that this is a
claim about reweighted theories only, with no reference to the
underlying staggered theory.  In the final step, the rooted staggered theory is
reconstructed with the help of Eq.~(\ref{resum}), and the scaling of $\D_n$
in the rooted reweighted theory ensures that the rooted staggered theory
has the desired continuum limit.

Finally, we address a more practical issue.
We have argued that QCD is correctly described by
the observables of the coarse-lattice (rooted, staggered) theory
in the limit of infinitely many blocking steps.
Thanks to the pull-back mapping every coarse-lattice
observable is at the same time a fine-lattice observable and, as such, it can
in principle be computed directly on the (rooted) staggered ensemble.
This amounts to using smeared staggered fermion sources constructed
in a particular way from all the blocking kernels.
The question is, are we allowed to use any other (local) fine-lattice
interpolating fields, as is done in practice?
Normally, one proves that once renormalization factors have
been correctly taken into account, different interpolating fields
must give rise to the same physical predictions.
But once again we face the difficulty that, because of the non-locality,
it may be dangerous to rely on rooted staggered perturbation theory.
We believe that, here too,
the resolution is to rely on
reweighted perturbation theory.  By its very construction, the reweighted
theory ``knows'' about fine- and coarse-lattice fields alike,
and, therefore, it is capable of comparing coarse-lattice
and fine-lattice interpolating fields.  We expect that the usual
statements about independence of the physical predictions of
any particular choice of the interpolating fields will remain true
once this comparison is carried out.  This can in principle be
tested in perturbation theory.

\section{Staggered chiral perturbation theory}

Here we discuss point 3 of the Introduction, namely the low-energy effective
theory for staggered quarks. By definition, such a theory must include
discretization effects, in particular
taste-violations, although we expect the theory will go over into continuum
chiral perturbation theory (\chpt) in the $a\to0$ limit.
When the underlying staggered action
is unrooted, finding the effective theory is straightforward \cite{LEE-SHARPE,AUBIN-BERNARD}.
We call the result ``unrooted  staggered chiral perturbation theory,'' or u\schpt.
On the other hand, the effective theory must be non-trivial
in the presence of rooting,
since we know from Sec.~2 that non-localities are present, and that their effects
show up in the pion sector.

In Ref.~\cite{AUBIN-BERNARD}, it was proposed that the rooting could be taken
into account at the chiral level by locating the presence
of sea-quark loops in the meson diagrams
of u\schpt\ and multiplying each by $1/4$.  These loops can be found by
quark-flow arguments \cite{QUARK-FLOW}, but for present purposes it is more useful
to use the replica trick \cite{REPLICA}, which was already employed in Sec.~3.3
for weak-coupling perturbation theory.
 This version of staggered chiral perturbation
theory will be referred to as r\schpt. We emphasize that the ``r'' in r\schpt\ stands,
in the first instance, for ``replica.''  The rules of r\schpt,
described in more detail below, give a well defined procedure
for computing chiral amplitudes. The question is, however, whether r\schpt\ is
in fact the proper low-energy theory for rooted staggered quarks.  We will argue
that it is correct; in other other words, we argue that the ``r'' in r\schpt\ also
stands for ``rooted.''

Finding the correct chiral theory is important for several reasons.  The discussion in Sec.~2
indicates that the potential problems from rooting
show up in the IR, because of
the interplay of the physical ($m_\pi^{GB}$) and unphysical ($a\Lambda^2_{QCD}$)
IR scales.\footnote{
In the RG analysis of Sec.~3, it is plausible that $\e_n$ (Eq.~(\ref{eps}))
over-estimates the relative size of successive
terms in the expansion of the logarithm in Eq.~(\ref{resum}),
and that the relative size is actually
$a\Lambda^2_{QCD}/ m_r \sim a\Lambda^3_{QCD}/ (m_\pi^{GB})^2$.
This  ratio indeed contains the two IR scales $m_\pi^{GB}$ and $a\Lambda_{QCD}^2$.
}
Since the chiral theory
describes the most IR part of the theory, it provides
a laboratory for studying the effects of non-locality and seeing (one hopes)
how they go away as $a\to0$.  Indeed, as mentioned in Sec.~2, r\schpt\ already shows
unitarity violations at $a\not=0$ \cite{PRELOVSEK,cdt,cb}.
Moreover, even if
the validity of the rooting procedure were rigorously established at the quark level,
the chiral theory would still be crucial for controlling
the chiral and continuum extrapolation of simulation results \cite{MILC}. Since the
effects of taste violations are significant at lattice spacings available currently and
in the foreseeable future, we cannot do without a chiral theory that
takes such effects into account.
Finally, the chiral theory provides a non-perturbative handle
for relating the unrooted  staggered valence sector to the
rooted staggered sea sector.  In particular, if r\schpt\ is indeed correct,
it can be used to show that even though the valence sector
is unrooted, the theory does not behave as a ``mixed'' theory in which
valence and sea quarks have different lattice actions.

The u\schpt\ theory for a single unrooted staggered field (\ie, one flavor) is
derived through $\cO(a^2)$ by Lee and Sharpe \cite{LEE-SHARPE}. They start
by using the staggered symmetries to find the Symanzik action through $\cO(a^2)$.
Spurions can then be introduced for all terms that violate the
taste and chiral symmetries, and the chiral theory follows. The generalization
of this procedure to multiple flavors appears in Ref~\cite{AUBIN-BERNARD}. In the
unrooted case, the generalization is straightforward.
The proposal to take into account the fourth root, is, in the replica
approach, the following:

\vspace{-0.35cm}

\begin{itemize}
\item Replicate the sea-quark degrees of freedom at the chiral level,
replacing each by $n_R$ identical copies, where $n_R$ is a positive integer.
All $n_R$ copies get the same mass, and, if relevant, the same
source terms.
\vspace{-0.35cm}

\item Calculate order by order in the resulting u\schpt, keeping the $n_R$
dependence explicit.
\vspace{-0.35cm}

\item Replace $n_R$ by $1/4$ at the end.
\end{itemize}

\vspace{-0.35cm}

{\noindent Note that the  dependence on $n_R$ is completely determined at any finite order in u\schpt, so replacing
$n_R$ by $1/4$ is a well-defined procedure.\footnote{In Ref.~\cite{cb}, it was claimed that
the dependence on $n_R$ is strictly polynomial. That is in fact an oversimplification, since we would
like to sum the geometric series of hairpin diagrams to all orders, which introduces factors
of $n_R$ into the masses of flavor-neutral mesons, and hence into the denominators
of Feynman integrands. But the key point remains: the functional form of the $n_R$
dependence is completely determined at any finite order in u\schpt.}
As always in a chiral theory, we treat the low-energy constants (LECs) as free parameters for each $n_R$;
we are just trying to find the dependence of physical quantities on the LECs in the rooted staggered theory.}

The argument \cite{cb} that r\schpt\ is the correct chiral theory for
rooted staggered quarks stays completely within the context of chiral theories.
As emphasized by Sharpe \cite{sharpe}, this is possible because the
unfamiliar unphysical features of the rooted theory can be connected to
the much more familiar, but still unphysical,  features of a partially
quenched (and unrooted) theory.  While \PQchpt, the chiral theory in the
partially quenched case \cite{PQ}, is not on as
firm a theoretical footing as is ordinary \chpt\ for full (unquenched) theories \cite{Weinberg},
we have a large body of numerical evidence that \PQchpt\ (and the completely
quenched version \Qchpt\ \cite{QCHPT,QUARK-FLOW}) are in fact the correct chiral
descriptions of the corresponding lattice theories. The evidence
comes from Wilson, domain wall, and (for the quenched case) overlap quarks, as well as
staggered simulations.  Furthermore,  \PQchpt\ will be subject to even more
stringent numerical tests in the future.  So connecting the rooted case
to the partially quenched case is a useful step forward.

Reference~\cite{cb} starts by noting that
we know (trivially) how the fourth root works when there are four
degenerate flavors ($n_F\!=\!4$).  Since the fourth power of the
fourth root reproduces the determinant, the rooted $n_F\!=\!4$ theory
is identical to the unrooted $n_F\!=\!1$ theory. We therefore know the starting
chiral theory, namely the u\schpt\ of Lee and Sharpe.
To get to a non-degenerate 4-flavor theory, we can
expand around the degenerate, massive point; this is where partial quenching
is needed. Finally,
to get to a theory with $n_F<4$ flavors, one  quark at a time can be decoupled.
Each step in the procedure requires some assumptions, which are plausible
but unproven.  However, most of the assumptions can be tested numerically,
and some already have been tested.

To explain the argument in more detail we need some notation.
Let
$(n_F,n_T,n_R)_{LQCD}$ be the generating functional for a lattice QCD theory with
$n_F$ flavors, $n_T$ tastes, and $n_R$ replicas of each flavor; let
$(n_F,n_T,n_R)_{\chi}$ be the generating functional of the corresponding chiral theory.
Whenever $n_R$ is shown explicitly it is taken to be a positive integer; the end result
of the replica trick is indicated by replacing $n_R$ with $1/4$.
When $n_R$ is trivially equal to 1 (because the replica trick is not relevant), it is omitted.
An unrooted theory is indicated by $n_T=4$; while for the rooted theory we put $n_T=1$.

Thus
$(1,4)_{LQCD}$ is the theory of a
single unrooted staggered field, and
$(1,4)_{\chi}$ is the u\schpt\ of Ref.~\cite{LEE-SHARPE}.
Similarly, $(n_F,4,n_R)_{\chi}$ is the u\schpt\ of Ref.~\cite{AUBIN-BERNARD}
with $n_R\!\cdot\! n_F$ sea-quark species.
Further, $(n_F,1)_{LQCD}$ is the lattice  theory of $n_F$ rooted staggered fields, and $(n_F,1)_{\chi}$
is by definition the low-energy theory
generated by  integrating out all the higher modes in $(n_F,1)_{LQCD}$.
As far as we know so far, $(n_F,1)_{\chi}$ could be horribly
non-local, non-unitary, or otherwise sick.  The claim, however, is that
$(n_F,1)_{\chi}$
is in fact $(n_F,4,\fourth)_\chi$ to any finite order in chiral
perturbation theory, where the latter simply
defines what we mean by r\schpt.

The first steps of the argument are taken in the $n_F\!=\!4$ case. We want to show that
\begin{equation}\eqn{toshow}
(4,1)_\chi \doteq (4,4,\fourth)_\chi \ ,
\end{equation}
where ``$\doteq$'' is used to indicate that the
two sides are the same functions of the LECs.
As mentioned above, we start with a degenerate $n_F\!=\!4$ theory, with mass matrix  $\cM = \bar m I$, where $I$ is
the identity matrix in flavor and taste space. In such as theory, we have
\begin{eqnarray}
\eqn{deg-root-lqcd}
(4,1)_{LQCD}\Big\vert_{\cM=\bar m I} & = & (1,4)_{LQCD}\Big\vert_{\bar m} \\
\eqn{deg-root-chi}
(4,1)_{\chi}\Big\vert_{\cM=\bar m I} & \doteq & (1,4)_{\chi}\Big\vert_{\bar m} \ \doteq\  (4,4,\fourth)_\chi\Big\vert_{\cM=\bar m I} \ .
\end{eqnarray}
The last equivalence here is manifest order by order in r\schpt, since taking
$4n_R$ degenerate flavors and then putting $n_R=1/4$ gives the same chiral expansion
as in a one-flavor theory.

To move away from degenerate limit, we add taste-singlet scalar sources $s^{ij}$ for the sea-quark fields:
\begin{eqnarray}
\cL_{(4,1)} &=& \dots+ \bar m\, \bar \Psi_i(x) \Psi_i(x) + \bar \Psi_i(x) \, s^{ij}(x)\,
\Psi_j(x) + \dots \nonumber \\
\cL_{(4,4,n_R)} &=& \dots+ \bar m \bar \Psi^r_i(x) \Psi^r_i(x) + \bar \Psi^r_i(x) \,s^{ij}(x)
\, \Psi^r_j(x)  + \dots \eqn{sea-sources} \ ,
\end{eqnarray}
with implicit sums over the flavor indices $i,j$ and the replica index $r$.
In the $(4,1)_{LQCD}$ theory the fourth root is taken {\it after}\/ the above
sources have been included in the determinant.

When $s\not=0$, $(4,4,\fourth)_\chi$ might not be the right chiral theory,
so we define the mismatch by
\begin{equation}\eqn{mismatch}
(4,1;s)_\chi \doteq (4,4,\fourth;s)_\chi + V[s] \ .
\end{equation}
The mismatch $V[s]$ can be expanded around $s=0$,
where it is known to vanish.
In fact, it can be shown \cite{cb} that all derivatives of
$(4,1;s)_\chi$ and $(4,4,\fourth;s)_\chi$ with respect
to the sources are equal:
\begin{equation}\eqn{derivs}
\prod_n\frac{\partial}{\partial s^{i_nj_n}(x_n)}
(4,1;s)_\chi\Big\vert_{s=0}\doteq \;\;
\prod_n\frac{\partial}{\partial s^{i_nj_n}(x_n)}
(4,4,\fourth;s)_\chi\Big\vert_{s=0} \ ,
\end{equation}
which implies
\begin{equation}\eqn{Vs-derivs}
\hspace{0.5cm} \prod_n\left(\frac{\partial}{\partial s^{i_nj_n}(x_n)}
 V[s]\right)\Bigg\vert_{s=0} = 0\ .
\end{equation}
\Equation{derivs} is proved by first relating sea-quark correlation functions in each theory
to partially quenched valence-quark correlation functions, where the valence quarks
have the same lattice action as the sea quarks.
This then allows us to set the sea-quark source $s=0$, where the equivalence of
the two chiral theories is known.
The proof requires the existence of standard \PQchpt\ for
unrooted theories, namely for $(1,4)_\chi$
and $(4,4,n_R)_\chi$.
We note that partially quenched theories are crucially needed because the derivatives with
respect to sea-quark sources in a rooted theory result in different factors
of $1/4$ in different contractions, which can never happen in the  sea-quark sector
of an unrooted theory.

Thus all derivatives of $V[s]$ vanish at $s=0$.
If $V[s]$ is analytic in $s$ --- up to possible isolated singularities ---
it must vanish everywhere, and
$(4,1;s)_\chi \doteq (4,4,\fourth;s)_\chi$.
Since an arbitrary four-flavor mass
matrix can be obtained by a suitable choice of $s$,
\eq{toshow} is verified.   In other words,
r\schpt$\;$  (\ie, $(4,4,\fourth)_\chi$) is the right chiral theory
in the four-flavor case.

The assumption of analyticity could fail in a two ways.
First of all, there could be a phase transition when the masses
are some finite distance away from the degenerate point.  There
is some evidence from simulations \cite{MILC} that this does not
occur, although it is still possible that a phase transition
lurks beyond the range of parameters (masses, lattice spacings)
that have been investigated to date.  Analyticity could also fail
if there were an essential singularity right at $s=0$.  This is
hard to rule out {\it a priori}, though it seems unlikely, since
we are expanding around a massive theory with no obvious IR
divergences.  At the end of
this section, we discuss work in progress on deriving r\schpt\ directly
from the lattice theory, using the methods of Sec.~3.
If successful, this work would (among other advantages) reduce
these analyticity concerns.

To move from $n_F\!=\!4$ to
$n_F\!=\!3$, we take one quark mass large. Call it the
``charm'' quark, with mass $m_c$.
We first choose $m_c$ as large as possible without leaving the region where chiral
perturbation theory applies. This point is taken,
nominally, as $m_c\sim  2m_s^{\rm phys}$, where
$m_s^{\rm phys}$ is the physical strange quark mass. For a clean
separation of scales, we can
temporarily take the other three masses much smaller than $m_s^{\rm phys}$.

We then integrate out $m_c$ from $(4,4,\fourth)_\chi$.
Since this is a perturbative process (order by order in r\schpt), there is little doubt
that the resulting chiral theory is $(3,4,\fourth)_\chi$, just as continuum $SU(2)_L\times
SU(2)_R\;$ \chpt\ results from integrating out the strange quark from the
$SU(3)_L\times SU(3)_R$ theory \cite{GASSER-LEUTWYLER}. Nevertheless, an
explicit check of this step is in progress \cite{BERNARD-DU}.

Since, by the previous steps in the argument,
$(4,4,\fourth)_\chi$ describes the long-distance physics of the $n_F=4$ theory,
$(3,4,\fourth)_\chi$  describes that physics when
$m_c\sim  2m_s^{\rm phys}$. We now {\it assume}\/ that this decoupling of the charm quark
from the long-distance physics remains true
(up to the usual renormalizations of relevant and marginal operators)
as $m_c$ increases still further, until $m_c \gg 1/a$.
At that point, $m_c$ is much larger than all the eigenvalues of the Dirac operator,
and charm must decouple from the lattice theory, leaving $(3,4,\fourth)_{LQCD}$.
Under our assumptions, we then have
\begin{equation}\eqn{nf3-result}
(3,1)_\chi \doteq  (3,4,\fourth)_{\chi} \ .
\end{equation}
So r\schpt\ is the correct chiral theory for three rooted staggered flavors.
We can then repeat the steps to argue
$(2,1)_\chi \doteq  (2,4,\fourth)_{\chi}$,
and $(1,1)_\chi \doteq  (1,4,\fourth)_{\chi}$.

These results immediately suggest an apparent paradox, which is
seen most clearly in the $n_F\!=\!1$ case.
The theory with one flavor should have only a heavy pseudoscalar, which we call
the $\eta'$, and no light pseudo-Goldstone bosons.
Yet the theory with one rooted staggered flavor
contains light pions, whose masses vanish in the continuum chiral limit, as well as one heavy meson,
the taste-singlet $\eta'_I$. There are different weightings of the contributions of such particles
in the chiral theories of the rooted and unrooted cases, but otherwise r\schpt\ and u\schpt\ are
similar.

We first note that the contributions
of the unwanted particles must disappear from
physical correlation functions in the continuum limit,
either by decoupling, or by canceling against each other, or both.
This follows from the argument  around Eq.~(\ref{cl}): In the continuum
limit taste symmetry is exact, and the physical sector (defined here as the correlation
functions generated by taste-singlet sources) is exactly equivalent to a continuum
theory of a single flavor. What r\schpt\ adds to the discussion is that it allows
us to see exactly how the decouplings and cancellations take place as the continuum
limit is approached.

An r\schpt\ calculation of the $n_F=1$ taste-singlet scalar-scalar correlator
has been performed to one loop \cite{cdt,cb}.  There are
two kinds of unphysical contributions that appear at $a\not=0$.  Terms
coming from  the taste-violating hairpins \cite{AUBIN-BERNARD}
involve taste-vector and taste-axial-vector pions, and are proportional to
explicit powers of $a^2$ --- they simply vanish in the continuum limit.
There are also contributions from the physical, taste-singlet
hairpins due to the anomaly, as well as from connected (non-hairpin) meson correlators.
These generate a two-$\eta'_I$ intermediate state with the physical weight,
but also two-pion states.  When we let $n_R\to1/4$, the latter
states have relative weights: 1, 4, 6, 4, -15, for the
taste-pseudoscalar (Goldstone), axial, tensor, vector, and
singlet pions, respectively.\footnote{See Sec.~VI of Ref.~\cite{cb} for
a detailed discussion of how these weights arise in r\schpt.}
The existence of the term with negative weight
here is a clear indication of unitarity violations.
At $a\not=0$, these pions have different masses due
to taste-symmetry violation and have a non-zero contribution to the
correlation function.  The sum of the weights is 0, however,
so the contributions cancel in the continuum limit when all the pions become degenerate.

If, as argued,
r\schpt\ is the correct chiral theory for rooted staggered quarks,
then there are some important consequences.
When $a\!\to\!0$, $(n_F,4,n_R)_{\chi}$ becomes ordinary \chpt\
for $4n_F\cdot n_R$ quark species.
Therefore, taking $n_R\to1/4$ order by order
produces standard, continuum \chpt\ for $n_F$ flavors in the physical
sector.\footnote{We
are considering here the case of positive quark masses only.  For
a discussion of the issues involved with negative masses,
see Refs.~\cite{bgss,Durr:2006ze}.} This then
implies that the lowest-energy regime of $n_F$-flavor lattice
QCD with rooted staggered quarks becomes
indistinguishable in structure in the continuum limit
from ordinary $n_F$-flavor QCD.   In this limit, there will therefore be
no unitarity or locality violations in the physical sector of the chiral theory.
Of course, assuming the arguments of Sec.~3 go through,  this {\it had}\/ to work, since
the continuum limit of the rooted staggered lattice theory is true QCD.

Another use of r\schpt\ is more technical.
It has been suggested (\eg in Ref~\cite{KENNEDY}) that the theory of rooted staggered
sea quarks  and (necessarily unrooted) staggered valence quarks
is a ``mixed'' theory, which behaves like it has different lattice actions
for the valence and sea sectors.  If that were the case, it would be rather unpleasant:
Among other features, mixed theories have different
renormalizations of sea and valence quark masses, and even if the quark masses
are tuned to make the sea-sea and valence-valence mesons degenerate,
the valence-sea mesons will be split from the others by discretization errors \cite{MIXED}.

In weak-coupling perturbation theory, it is easy to see with the replica
trick that the rooted staggered theory does not behave like a mixed
theory: the renormalizations in the  sea and valence
sectors are the same \cite{cb,ys}. Furthermore, r\schpt\ allows us to argue that
the theory is not mixed at the chiral level either.
The point is that $(n_F,4,\fourth)_\chi$ is obtained order by order from $(n_F,4,n_R)_\chi$.
The latter theory in turn has lattice symmetries (broken only by mass terms)
that interchange valence and sea quarks.
These symmetries imply that the theory of rooted staggered sea quarks and
staggered valence quarks behaves like a partially quenched theory, not a mixed theory.
One can give different masses to the sea and valence quarks,
as usual in a partially quenched situation.  However, if one chooses to give
valence and sea quarks the same masses, then the symmetries forbid, for example, the splitting
of valence-sea from valence-valence or sea-sea mesons.

Finally, we briefly describe our
work in progress \cite{CBMGYS}, in which we attempt to
derive r\schpt\ directly from the rooted staggered lattice theory.
The immediate problem in such a derivation is that
a straightforward replica trick fails at the non-perturbative
QCD level because we have no control over the functional
dependence of the theory on $n_R$. In weak-coupling QCD perturbation theory the
dependence on $n_R$ arises only from the counting of sea quark
loops, and thus is manifest order by order.  But once we move
beyond perturbation theory, the $n_R$ dependence is not known
{\it a priori}\/, and a unique analytic continuation from integer $n_R$ to
$n_R=1/4$ is not possible.

This problem with the replica trick at the QCD level
is to be contrasted with the trick at the chiral level: r\schpt\
makes sense because the dependence on $n_R$ is known
when we calculate order by order in chiral perturbation
theory.  The unknown $n_R$ dependence of the QCD level is hidden
in the $n_R$ dependence of the LECs of the corresponding chiral theory.
Since the goal in the chiral theory is simply to calculate physical
quantities as functions of the LECs, which are treated as independent
variables, we can (must!) ignore the hidden $n_R$ dependence of the
LECs in r\schpt.

Our tentative solution to this problem is to use the reweighted theory
of Sec.~3 as an intermediate step.
Since this is a local theory, finding the corresponding
chiral theory is straightforward. We then need the replica trick
only to move from the reweighted theory to the
rooted staggered case.  The latter can be obtained by a convergent Taylor
expansion (in the parameter $t$ introduced after Eq.~(\ref{splitn})) around the former.
This makes it possible to control the $n_R$ dependence at the QCD level, which
should allow us to establish an unambiguous
connection between the QCD and chiral levels, and thereby to derive r\schpt\ directly.
It should be noted, however, that such a derivation would still rely on the
existence of \PQchpt\ for standard (local) partially quenched theories.

\section*{Acknowledgments}

We thank Steve Sharpe and Andreas Kronfeld for useful discussions.
All three authors thank the Institute for Nuclear Theory at the University
of Washington, Seattle, for hospitality; CB also thanks
the Centre de Physique Th\'eorique, Marseille.  MG was supported in part by the
Generalitat de Catalunya under program PIV1-2005.
CB and MG are supported in part by the US Department of Energy, and
YS is supported by the Israel Science Foundation under grant
222/02-1.

\end{document}